\documentclass{article}
\usepackage{times}
\usepackage{amsfonts}
\usepackage{graphicx}
\begin{document}
\noindent
{\Large  BOLTZMANN ENTROPY OF A NEWTONIAN UNIVERSE}
\noindent

\vskip.5cm
\noindent
{\bf D. Cabrera},$^{a}$ {\bf P. Fern\'andez de C\'ordoba},$^{b}$ and {\bf  J.M. Isidro}$^{c}$\\
Instituto Universitario de Matem\'atica Pura y Aplicada,\\ Universidad Polit\'ecnica de Valencia, Valencia 46022, Spain\\
${}^{a}${\tt dacabur@upvnet.upv.es}, ${}^{b}${\tt pfernandez@mat.upv.es}\\
${}^{c}${\tt joissan@mat.upv.es}\\

\vskip.5cm
\noindent
{\bf Abstract}  A dynamical estimate is given for the Boltzmann entropy of the Universe, under the simplifying assumptions provided by Newtonian cosmology. We first model the cosmological fluid as the probability fluid of a quantum--mechanical system. Next, following current ideas about the emergence of spacetime, we regard gravitational equipotentials as isoentropic surfaces. Therefore gravitational entropy is proportional to the vacuum expectation value of the gravitational potential in a certain quantum state describing the matter contents of the Universe. The entropy of the matter sector can also be computed. While providing values of the entropy that turn out to be somewhat higher than existing estimates, our results are in perfect compliance with the upper bound set by the holographic principle.

\tableofcontents

\section{The approach via emergent quantum theory}\label{habel}

In this article we will argue in favour of {\it emergent}\/ quantum mechanics as providing an appropriate framework to estimate thermodynamical quantities  of the Universe, such as the entropy. 

The notion that {\it quantum mechanics is an emergent theory}\/ has been discussed at length in the literature  \cite{ADLER, TOPOLOGICAL, COMPLEXANALYTIC, ELZE, NAVIER, GALLEGO, KHRENNIKOV}. Combined with the idea that {\it spacetime as well is an emergent phenomenon}\/ \cite{PADDY1, PADDY3,  PADDY4, PADDY5, VERLINDE}, this paves the way for {\it a computation of some thermodynamical properties of spacetime in quantum--mechanical terms}\/. We would like to remark that a quantum--mechanical approach to the expansion of the Universe was called for long ago in ref. \cite{EDDINGTON}, where it was suggested to regard the expansion of the Universe as a scattering problem in quantum mechanics. 

The expansion of the Universe is a long--standing experimental observation \cite{HUBBLE}, that has in more recent years been refined thanks to very precise measurements \cite{PERLMUTTER, RIESS}. In the Newtonian approximation, this receding behaviour of the galaxies can be easily modelled  by a phenomenological potential, namely, an isotropic harmonic potential carrying a negative sign: 
\begin{equation}
U_{\rm Hubble}({\bf r})=-\frac{H_0^2}{2}{\bf r}^2. 
\label{potenzi}
\end{equation}
As the angular frequency  we take the current value of Hubble's constant $H_0$. (Thus $U_{\rm Hubble}$ has the dimensions of energy per unit mass, or velocity squared).

In the emergent approach to spacetime presented in ref. \cite{VERLINDE}, gravity qualifies as an entropic force. If gravitational forces are entropy gradients, gravitational equipotential surfaces can be identified with isoentropic surfaces. Recalling the arguments of ref. \cite{VERLINDE}, a classical point particle approaching a holographic screen causes the entropy of the latter to increase by one quantum $k_B$. Here we will analyse a quantum--mechanical model in which the forces driving the galaxies away from each other can be modelled by the Hubble potential (\ref{potenzi}).  We will replace the classical particle of ref. \cite{VERLINDE} with a collection of quantum particles (the matter contents of the Universe) described by the wavefunction $\psi$. Let $U$ denote the gravitational potential. Once dimensions are corrected (using $\hbar$ and  $k_B$), the expectation value $\langle\psi\vert U\vert\psi\rangle$ becomes the quantum--mechanical analogue of the entropy increase caused by a classical particle approaching a holographic screen. Thus {\it the expectation value $\langle\psi\vert U\vert\psi\rangle$ is a measure of the gravitational entropy of the Universe when the matter  of the Universe is described by the wavefunction $\psi$}\/.  

The potential $U_{\rm Hubble}$ of Eq. (\ref{potenzi}) encodes the combined effect of the gravitational attraction, and of the repulsion caused by the dark energy on the matter content of the Universe (baryonic and dark matter).  We can therefore identify the Hubble potential $U_{\rm Hubble}$ of Eq. (\ref{potenzi}) with the gravitational potential $U$ in the previous paragraph. Let us briefly recall  why $U_{\rm Hubble}$ in fact combines a Newtonian gravitational attraction, plus a harmonic repulsion. In the Newtonian limit considered throughout in this paper, the gravitational attraction is computed by applying Gauss' law to a sphere filled with a homogeneous, isotropic density of matter. Then the gravitational field {\it within the sphere}\/ turns out to be proportional to the position vector, so the corresponding potential becomes a quadratic function of the position. Altogether, the total potential at any point within the cosmological fluid is the sum of two harmonic potentials; Hubble's constant $H_0$ is the frequency of this total harmonic potential.

We will first start with a flat, Euclidean space governed by nonrelativistic, Newtonian cosmology. The latter  leads to  the same equation that governs the scale factor of general--relativistic Robertson--Walker models \cite{WEINBERG}. On the other hand it circumvents the mathematical sophistication required by general relativity. The advantage of first performing a nonrelativistic treatment is that it bears out the deep connection existing between the cosmological fluid, and the Madelung approach to Schroedinger quantum mechanics; this is done in section \ref{shache}. In section \ref{hala}  we obtain a perturbative estimate for the entropy of the Universe; finally this analysis is carried out nonperturbatively in section \ref{npetrp}. As a next level of sophistication we move on to a flat Friedmann--Lema\^\i tre--Robertson--Walker (FLRW) 4--dimensional spacetime; due to the great length of the calculations involved, the corresponding results will be presented in an upcoming publication \cite{UPCOMING}. 

We would finally like to stress the following points:

{\it i)} The wavefunction $\psi$ we will be concerned with here is meant to provide a {\it phenomenological}\/ description of the receding matter  in its recessional motion {\it within a fixed spacetime}\/.

{\it ii)} We will comply with the cosmological principle, the latter stated in either one of the following two (inequivalent) ways. In its first formulation, given the wave function $\psi$, the volume density of matter $\vert\psi\vert^2$ is spatially constant. In its second formulation, given $\psi$ and the 3--dimensional volume element ${\rm d}^3V=\sqrt{\vert g\vert}\,{\rm d}x^1{\rm d}x^2{\rm d}x^3$, the particle number $\vert\psi\vert^2{\rm d}^3V$ is spatially constant. This second formulation is weaker than, and implied by, the first one.

{\it iii)} A quantum--mechanical wavefunction $\psi$ for the matter contents of the Universe will be used to obtain an estimate of the gravitational entropy of the Universe. 

{\it iv)} Invoking Boltzmann's principle, the same wavefunction $\psi$ can be used to obtain an estimate of the entropy of the matter of the Universe (baryonic and dark matter).

{\it v)} All entropies referred to in this paper are Boltzmann entropies.

We would like to thank the referees for drawing our attention to a number of issues and papers, where points related to those analysed here are dealt with (from different perspectives). Specifically: a wavefunction of the Universe was first considered in ref. \cite{DEWITT}; the quantum potential was shown to generate gravitational attraction between particles in ref. \cite{MATONEGRAVITY}; the Hubble potential possesses no groundstate when considered on all of $\mathbb{R}^3$ \cite{BROADBRIDGE1}; the inexistence of a stable groundstate has consequences on quantum fields on an expanding spacetime \cite{BROADBRIDGE2}. Since we are considering the Hubble potential on a finite Universe (with radius $R_0$), the existence of a stable groundstate is guaranteed. Also, the Hubble expansion is considered to maintain the Universe close to equilibrium, so we can apply standard thermodynamical relations.

\section{Newtonian cosmology \`a la Madelung}\label{sha}

Newtonian cosmology represents a first step that, while avoiding the technical difficulties of general relativity, succeeds in capturing some essential physics of the Universe \cite{MEXICO, WEINBERG}.

\subsection{The ideal--fluid description} \label{shache}

In this section we will establish that Newtonian cosmology can be conveniently regarded as a nonrelativistic quantum mechanics. 

In Newtonian cosmology, the matter content of the Universe is modelled as an ideal fluid satisfying the continuity equation and the Euler equation,
\begin{equation}
\frac{\partial\rho}{\partial t}+\nabla\cdot\left(\rho{\bf v}\right)=0,\qquad \frac{\partial{\bf v}}{\partial t}+\left({\bf v}\cdot\nabla\right){\bf v}+\frac{1}{\rho}\nabla p-{\bf F}=0.
\label{knoott}
\end{equation}
Above, $\rho$ is the volume density of fluid mass, $p$ is the pressure, ${\bf v}$ is the velocity field, and ${\bf F}$ the force per unit volume acting on the fluid. The cosmological principle requires $\rho$ and $p$ to be spatially constant; it also leads to the requirement that the velocity ${\bf v}$ be everywhere proportional to the position vector ${\bf r}$. This latter requirement is nothing but Hubble's law, so the Hubble potential (\ref{potenzi}) arises naturally as a consequence of the cosmological principle. The gravitational self--attraction of the matter distribution, plus Hubble's repulsion, are both taken care of by the force ${\bf F}$ in the Euler equation.

Madelung long ago reexpressed Schroedinger quantum mechanics also in terms of an ideal fluid. Specifically, one separates the nonrelativistic wavefunction $\psi$ into amplitude and phase,
\begin{equation}
\psi=A\exp\left(\frac{{\rm i}}{\hbar}{\cal I}\right)=\exp\left(S+\frac{{\rm i}}{\hbar}{\cal I}\right), \qquad A=:{\rm e}^S=\exp\left(\frac{{\cal S}}{2k_B}\right),
\label{benouere}
\end{equation}
(Here ${\cal I}$ is the classical mechanical action integral; we will later on invoke Boltzmann's principle to regard  ${\cal S}$ as the Boltzmann entropy and $S:={\cal S}/2k_B$ as the dimensionless Boltzmann entropy). Substituting the Ansatz (\ref{benouere}) into the Schroedinger equation for $\psi$, one is led to an expression containing a real part and an imaginary part. The imaginary part  turns out to be the continuity equation for the quantum probability fluid,
\begin{equation}
\frac{\partial S}{\partial t}+\frac{1}{m}\nabla S\cdot\nabla {\cal I}+\frac{1}{2m}\nabla^2{\cal I}=0,
\label{konntt}
\end{equation}
where the velocity field ${\bf v}$ and the density $\rho$ are defined by
\begin{equation}
{\bf v}:=\frac{1}{m}\nabla {\cal I}, \qquad \rho=A^2={\rm e}^{2S}.
\label{vanednie}
\end{equation}
The real part turns out to be the quantum Hamilton--Jacobi equation:
\begin{equation}
\frac{\partial {\cal I}}{\partial t}+\frac{1}{2m}(\nabla {\cal I})^2+V+{\cal Q}=0,
\label{vabueelnv}
\end{equation}
where $V=mU$ is the external potential, and we have introduced the quantum potential \cite{MATONE}
\begin{equation}
{\cal Q}:=-\frac{\hbar^2}{2m}\frac{\nabla^2 A}{A}.
\label{vebiyer}
\end{equation}
{}Finally, an Euler equation for this quantum probability fluid is obtained by taking the gradient of the quantum Hamilton--Jacobi Eq. (\ref{vabueelnv}):
\begin{equation}
\frac{\partial{\bf v}}{\partial t}+\left({\bf v}\cdot\nabla\right){\bf v}+\frac{1}{m}\nabla {\cal Q}+\frac{1}{m}\nabla V=0.
\label{kblm}
\end{equation}

We conclude that a 1--to--1correspondence between the cosmological fluid, on the one hand, and the quantum probability fluid, on the other, is provided by the following replacements:
\begin{equation}
\rho\mapsto{\rm e}^{2S}, \qquad {\bf v}\mapsto\frac{1}{m}\nabla {\cal I}, \qquad \frac{1}{\rho}\nabla p\mapsto\frac{1}{m}\nabla {\cal Q},
\qquad{\bf F}\mapsto-\frac{1}{m}\nabla V.
\label{korrer}
\end{equation} 
The above correspondence suggests that, {\it given the cosmological fluid in the Newtonian approximation, we use nonrelativistic quantum mechanics as an equivalent description thereof}\/. The value of $m$ entering Eq. (\ref{korrer}) is that of the overall matter contents of the Universe (baryonic and dark matter). This matter is subject to the repulsive effect of the dark energy, and to its own gravitational selfattraction, the combined effect of which is modelled by the effective Hubble potential  (\ref{potenzi}).

\subsection{Perturbative estimate of the entropy}\label{hala}

\subsubsection{Wavefunction of the matter distribution}

We will model the ideal fluid of Newtonian cosmology by means of the probability fluid corresponding to a {\it scalar field $\psi$ satisfying the Schroedinger equation}\/. Initially the latter will be taken to be the free equation, for a perturbative treatment. That is, $\psi$ will be used to compute $\langle\psi\vert U\vert\psi\rangle$, with $U$ the Hubble potential (\ref{potenzi}). Alternatively we can include the Hubble potential (\ref{potenzi}) in the Schroedinger equation already from the start; this nonperturbative treatment will be carried out in section \ref{npetrp}.

The squared modulus $\vert\psi\vert^2$ will equal the volume density $\rho$ of mass. The cosmological principle requires the density $\rho$ to be constant across space. In turn, the correspondence (\ref{korrer}) implies that $S$ must be a constant, so the quantum potential $Q$ will vanish identically. Again by the correspondence (\ref{korrer}), the pressure $p$ must be spatially constant, which is also in agreement with the cosmological principle. The effective Hubble potential will be introduced later on, as a perturbation to the free field $\psi$. 

The free Schroedinger equation admits the plane--wave solutions
\begin{equation}
\psi_{{\bf k}}({\bf r})=\frac{1}{R_0^{3/2}}\exp\left({\rm i}{\bf k}\cdot{\bf r}\right).
\label{wfkuu}
\end{equation}
They have been normalised within a cubic box of side $R_0$, the radius of the observable Universe. The cosmological principle is satisfied in its first formulation as given in section \ref{habel}. Moreover, the constant amplitude $A=R_0^{-3/2}$ leads to a vanishing quantum potential in (\ref{vebiyer}), in agreement with previous requirements.

The free Schroedinger equation can also be separated in spherical coordinates. The resulting free spherical waves $\psi_{\kappa lm}(r,\theta, \phi)$ are then labelled by $\kappa$ (the modulus of the linear momentum ${\bf k}$) and $l,m$ (the angular momentum quantum numbers). The cosmological principle imposes $l=0$. We will therefore consider the free spherical waves
\begin{equation}
\psi_{\kappa 00}(r,\theta,\phi)=\frac{1}{\sqrt{4\pi R_0}}\frac{1}{r}\exp\left({\rm i}\kappa r\right).
\label{iggn}
\end{equation}
They have been normalised within a sphere of radius $R_0$, instead of a cubic box. Once the spherical Jacobian factor $4\pi r^2$ is taken into account, the second formulation of the cosmological principle given in section \ref{habel} is satisfied. Moreover, the amplitude $A=1/r$ also leads to a vanishing quantum potential in (\ref{vebiyer}) because
\begin{equation}
\frac{\nabla^2 A}{A}=r\nabla^2\left(\frac{1}{r}\right)=-4\pi r\delta({\bf r})=0.
\label{vamnss}
\end{equation}

We will use both plane waves (\ref{wfkuu}) and spherical waves (\ref{iggn}) in order to model the distribution of the matter contents of the Universe. The results obtained from one or the other can at most differ by a dimensionless factor  of geometrical origin, due to the use of a cubic box as opposed to a spherical box. Imposing boundary conditions on the wavefunction at the walls of the corresponding box only leads to a quantisation of the energy levels; a possibility that we will disregard here (see section \ref{remaque} for a discussion of this point).

Since we are not imposing boundary conditions, we will work with a set of two linearly independent solutions to the free Schroedinger equation. In Cartesian coordinates, a fundamental set of solutions is provided by the wavefunctions $\psi_{\pm{\bf k}}(x,y,z)$; in spherical coordinates, a fundamental set of solutions is provided by the wavefunctions $\psi_{\pm\kappa 00}(r,\theta,\phi)$.

\subsubsection{Expectation values}

{}From what was said above, the operator ${\bf R}^2=X^2+Y^2+Z^2$, which is proportional to the effective potential (\ref{potenzi}), is a measure of the amount of gravitational entropy enclosed by the Universe.  Specifically, the combination
\begin{equation}
{\cal S}_{g}:={\cal N}\frac{k_B m H_0}{\hbar}{\bf R}^2
\label{nachhause}
\end{equation}
is dimensionally an entropy; a {\it dimensionless}\/ factor ${\cal N}$ is of course left undetermined. We call ${\cal S}_g$ the gravitational entropy operator. Its expectation value in the cubic--box state (\ref{wfkuu}) equals
\begin{equation}
\langle\psi_{\bf k}\vert{\cal S}_g\vert\psi_{\bf k}\rangle={\cal N}\frac{k_B m H_0}{\hbar}R_0^2,
\label{poopuy}
\end{equation}
while in the spherical--box state (\ref{iggn}) it reads
\begin{equation}
\langle\psi_{\kappa 00}\vert{\cal S}_g\vert\psi_{\kappa 00}\rangle={\cal N}\frac{k_BmH_0}{\hbar}\frac{R_0^2}{3}.
\label{vfertac}
\end{equation}
Substituting the known values of the cosmological data \cite{PLANCK} into Eqs. (\ref{poopuy}) and (\ref{vfertac}) we find the estimate
\begin{equation}
\frac{\langle {\cal S}_g\rangle}{k_B}\simeq 10^{123},
\label{estimado}
\end{equation}
where we have (arbitrarily) set ${\cal N}=1/2.6$ when using the plane--wave result (\ref{poopuy}), and ${\cal N}=3/ 2.6$ when using the spherical--wave result (\ref{vfertac}), in order to keep just powers of $10$. Our final result (\ref{estimado}) saturates the upper bound set by the holographic principle \cite{BOUSSO}.

We can also obtain an estimate for the entropy content of the matter described by the wavefunction $\psi$. Invoking Boltzmann's principle, one regards the amplitude $A$ of the wavefunction $\psi$ as the exponential of the entropy (in units of $k_B$) of the particles described by the wavefunction $\psi$.  This fact has been implicitly taken into account in the notation of Eq. (\ref{benouere}), from where we derive the entropy in terms of the amplitude:
\begin{equation}
{\cal S}_m=2k_B\ln A.
\label{piaintro}
\end{equation}
Acting on the plane waves (\ref{wfkuu}), the matter entropy operator ${\cal S}_m$ is a constant,
\begin{equation}
{\cal S}_m=-3k_B\ln R_0.
\label{piatte}
\end{equation}
Therefore its expectation value in the state (\ref{wfkuu}) equals
\begin{equation}
\langle\psi_{\bf k}\vert {\cal S}_m\vert\psi_{\bf k}\rangle=-3k_B\ln R_0.
\label{werrsd}
\end{equation}
The above is the correct behaviour for the entropy of an ideal gas, since the radius of the Universe is inversely proportional to its temperature. 

{}For the spherical waves (\ref{iggn}) we arrive at a matter entropy operator ${\cal S}_m$
\begin{equation}
{\cal S}_m=-2k_B\ln r-k_B\ln\left(4\pi R_0\right).
\label{enrottt}
\end{equation}
Its expectation value in the state (\ref{iggn}) is found to be
\begin{equation}
\langle\psi_{\kappa 00}\vert {\cal S}_m\vert\psi_{\kappa 00}\rangle=-3k_B \ln R_0,
\label{njvroer}
\end{equation}
after dropping a constant independent of $R_0$. We find again the expected (ideal--gas) logarithmic dependence of the entropy with respect to the temperature.

\subsection{Nonperturbative estimate of the entropy}\label{npetrp}

\subsubsection{Exact eigenfunctions}\label{vnasmn}

A nonperturbative evaluation requires solving the interacting Schroedinger equation $H\psi=E\psi$, where now
\begin{equation}
H=-\frac{\hbar^2}{2m}\nabla^2-\frac{k_{\rm eff}}{2}{\bf r}^2,\qquad k_{\rm eff}=mH_0^2.
\label{jami}
\end{equation}
Let us separate variables in Eq. (\ref{jami}) using spherical coordinates. The standard factorisation $\psi({\bf r})=R(r)Y_{lm}(\theta,\phi)$ leads to a radial wave equation
\begin{equation}
\frac{1}{r^2}\frac{{\rm d}}{{\rm d}r}\left(r^2\frac{{\rm d}R}{{\rm d}r}\right)-\frac{l(l+1)}{r^2}R+\frac{2m}{\hbar^2}\left(E+\frac{k_{\rm eff}}{2}r^2\right)R=0.
\label{radas}
\end{equation}
Two linearly independent solutions with $l=0$ are \cite{LEBEDEV}
\begin{equation} 
R_{\lambda}^{(1)}(r)=\exp\left(\frac{{\rm i}a^2}{2}r^2\right) {}_1F_1\left(\frac{3}{4}-\frac{{\rm i}\lambda}{4},\frac{3}{2}; -{\rm i}a^2r^2\right)
\label{herri}
\end{equation}
and
\begin{equation}
R_{\lambda}^{(2)}(r)=\frac{1}{r}\exp\left(\frac{{\rm i}a^2}{2}r^2\right){}_1F_1\left(\frac{1}{4}-\frac{{\rm i}\lambda}{4},\frac{1}{2}; -{\rm i}a^2r^2\right).
\label{tabernae}
\end{equation}
Above, ${}_1F_1(\alpha;\gamma;z)$ is the confluent hypergeometric function, and the parameters $a$, $\lambda$ can be expressed in terms of the mechanical data $m$, $k_{\rm eff}$, $E$, $H_0$:
\begin{equation}
a^4:=\frac{mk_{\rm eff}}{\hbar^2},\qquad \lambda:=\frac{2E}{\hbar H_0}.
\label{morcilla}
\end{equation}
The complete interacting wavefunctions are (up to radial normalisation factors)
\begin{equation}
\psi^{(j)}_{\lambda}(r,\theta,\phi)=\frac{1}{\sqrt{4\pi}}R_{\lambda}^{(j)}(r),\qquad j=1,2, \qquad\lambda\in\mathbb{R}.
\label{karkon}
\end{equation}
Since  $\lambda\in\mathbb{R}$ is the (dimensionless) energy eigenvalue, it plays the same role that the quantum number $n\in\mathbb{N}$ plays in the standard harmonic oscillator. Our harmonic potential does not have quantised energy levels, but continuous energy levels $\lambda$ instead. However the range of values covered by $\lambda$, while unbounded above, is bounded below by 
\begin{equation}
E_0=-\frac{1}{2}mH_0^2R_0^2
\label{petite}
\end{equation}
or, in terms of the dimensionless eigenvalue $\lambda$, by
\begin{equation}
\lambda_0=-\frac{mH_0R_0^2}{\hbar}=-2.6\times 10^{123}.
\label{teddy}
\end{equation}
Substituting this value of $\lambda_0$ into Eq. (\ref{karkon}) produces the wavefunctions $\psi_{\lambda_0}^{(j)}$, with $j=1,2$, which are the analogues of the vacuum wavefunction of the usual oscillator. The bound (\ref{petite}) has been determined by a purely classical argument; although the uncertainty principle will shift the minimum energy (\ref{petite}) by a positive amount, this correction can be discarded for our purposes, as it will be negligible compared to (\ref{petite}) itself.

As opposed to the free wavefunctions (\ref{wfkuu}) and (\ref{iggn}), the existence of zeroes of the confluent hypergeometric function ${}_1F_1$ is a sure sign that the cosmological principle will be violated by the wavefunctions (\ref{karkon}), but the extent of this violation remains to be determined. We claim that:\\
{\it i)} the expectation value of the quantum potential (\ref{vebiyer}) is a measure of the violation of the cosmological principle. More precisely, small values of the dimensionless ratio $\vert\langle{\cal Q}\rangle/\langle V\rangle\vert$ imply small violations of the cosmological principle, while large values imply large violations;\\
{\it ii)} the ratio $\vert\langle{\cal Q}\rangle/\langle V\rangle\vert$ achieves a minimum when evaluated in two states $\psi_{\lambda_0}^{(j)}$, because the numerator $\vert\langle{\cal Q}\rangle{\vert}$ reaches a minimum while  the denominator ${\vert}\langle V\rangle{\vert}$ reaches a maximum. That ${\vert}\langle V\rangle{\vert}$ achieves a maximum when $\lambda=\lambda_0$ is obvious. In what follows we would like to argue that ${\vert}\langle{\cal Q}\rangle{\vert}$ also reaches a minimum when $\lambda=\lambda_0$. 

Evaluating the quantum potential (\ref{vebiyer}) in terms of the eigenfunction $\psi$, with eigenvalue $E$, leads to
\begin{equation}
{\cal Q}=E-V+\frac{\hbar^2}{8m}\left[\psi^{-2}(\nabla\psi)^2+(\psi^*)^{-2}(\nabla\psi^*)^2-2(\psi^*\psi)^{-1}\nabla\psi^*\nabla\psi\right].
\label{quatitistiko}
\end{equation}
Its expectation value in the eigenstate $\psi$ equals
\begin{equation}
\langle{\cal Q}\rangle=E-\langle V\rangle+\frac{\hbar^2}{8m}\int\left[\psi^*\psi^{-1}(\nabla\psi)^2+ (\psi^*)^{-1}\psi(\nabla\psi^*)^2
-2\nabla\psi^*\nabla\psi\right].
\label{uchaspsi}
\end{equation}
Altogether, the ratio
\begin{equation}
\frac{\langle{\cal Q}\rangle}{\langle V\rangle}=\frac{E-\langle V\rangle}{\langle V\rangle}
+\frac{\hbar^2}{8m\langle V\rangle}\int \left[\psi^*\psi^{-1}(\nabla\psi)^2+(\psi^*)^{-1}\psi(\nabla\psi^*)^2-2\nabla\psi^*\nabla\psi\right] 
\label{provttbf}
\end{equation}
is a dimensionless number. If it vanishes, the eigenfunction $\psi$ satisfies the cosmological principle reasonably well. If the ratio (\ref{provttbf}) is nonvanishing but nevertheless small in absolute value, the eigenfunction $\psi$ will satisfy the cosmological principle at least approximately, and our computation of the entropy will be on a sound basis. 

Actually the ratio (\ref{provttbf}) depends on the energy eigenvalue $\lambda$. We expect a regime of values to exist for $\lambda$ such that, within this regime, the dimensionless ratio $\langle {\cal Q}\rangle/\langle V\rangle$ will be small enough to guarantee the validity of the replacement of the cosmological fluid with the quantum probability fluid. In order to justify this expectation we first observe that, for real eigenfunctions $\psi$, the ratio (\ref{provttbf}) simplifies considerably:
\begin{equation}
\frac{\langle{\cal Q}\rangle}{\langle V\rangle}=\frac{E-\langle V\rangle}{\langle V\rangle},\qquad \psi^*=\psi.
\label{semplifikato}
\end{equation}
Of course, our eigenfunctions (\ref{karkon}) are not real. However, still assuming $\psi^*=\psi$, the best possible ratio $\langle {\cal Q}\rangle/\langle V\rangle$ is attained for $E=\langle V\rangle$. This makes the following assumption plausible: {\it the complex wavefunction $\psi_0$ which minimises the ratio $\vert\langle {\cal Q}\rangle/\langle V\rangle\vert$ is that for which the energy eigenvalue $E_0$ equals the expectation value $\langle\psi_0\vert V\vert\psi_0\rangle$}\/. 

We therefore expect the two states $\psi_{\lambda_0}^{(j)}$ of Eq. (\ref{karkon}), with $\lambda_0$ given in Eq. (\ref{teddy}), to be those that {\it minimally}\/ violate the cosmological principle. In other words, the correspondence put forth in this paper (the quantum probability fluid as an equivalent description of the ideal cosmological fluid) works best when applied to the states $\psi_{\lambda_0}^{(j)}$, while progressively becoming less and less reliable as the energy increases.

Unfortunately, the exact vacuum--state eigenfunctions $\psi_{\lambda_0}^{(j)}$ of Eqs. (\ref{herri}) and (\ref{tabernae}) contain the huge parameter $\lambda_0$ of Eq. (\ref{teddy}). Due to the oscillatory behaviour of the confluent hypergeometric function, this renders the exact radial eigenfunctions (\ref{karkon}) extremely cumbersome to work with, both analytically and numerically. To simplify matters we will replace  the exact vacuum--state eigenfunctions $\psi_{\lambda_0}^{(j)}$ of Eqs. (\ref{herri}) and (\ref{tabernae}) with a set of approximate radial eigenfunctions for the vacuum state. We will also see that these approximate eigenfunctions will be real, so they will violate the cosmological principle only minimally.

\subsubsection{Approximate eigenfunctions for the vacuum state}\label{mashashas}

We set $l=0$ in Eq. (\ref{radas}) and use $E=E_0$ from (\ref{petite}) to arrive at the eigenvalue equation {\it for the vacuum state}\/:
\begin{equation}
\frac{1}{r^2}\frac{{\rm d}}{{\rm d}r}\left(r^2\frac{{\rm d}R}{{\rm d}r}\right)+\frac{m^2H_0^2}{\hbar^2}\left(r^2-R_0^2\right)R=0.
\label{taugcher}
\end{equation}
The change of variables
\begin{equation}
r=:R_0x, \qquad R(r)=:f(x),
\label{equis}
\end{equation}
where $x\in[0,1]$ is dimensionless, reduces Eq. (\ref{taugcher}) to
\begin{equation}
\frac{1}{x^2}\frac{{\rm d}}{{\rm d}x}\left(x^2\frac{{\rm d}f}{{\rm d}x}\right)+\sigma_0^2(x^2-1)f(x)=0, \qquad \sigma_0^2:=\frac{m^2H_0^2R_0^4}{\hbar^2}.
\label{ridotta}
\end{equation}
As compared to (\ref{taugcher}), the above equation is defined on the interval $x\in[0,1]$, which is more manageable than the original $r\in[0,R_0]$; moreover, all large numbers present in the problem are contained within the {\it dimensionless}\/ parameter $\sigma_0$ (the opposite of $\lambda_0$ in Eq. (\ref{teddy})):
\begin{equation}
\sigma_0=-\lambda_0=2.6\times 10^{123}.
\label{mundini}
\end{equation}
The parameter $\sigma_0$ equals the entropy of  Eq. (\ref{estimado}) in units of $k_B$; in fact, modulo the irrelevant factor $2.6$, it equals the holographic bound \cite{BOUSSO}. We conclude that {\it the radial wave equation (\ref{ridotta}) encodes information about the holographic principle}\/. 

We have seen in section \ref{vnasmn} that Eq. (\ref{ridotta}) is exactly soluble. However, the sheer size of $\sigma_0$ renders the exact wavefunctions (\ref{herri}) and (\ref{tabernae}) totally useless: analytical computations with them are out of the question, and numerical computations quickly get out of range. For this reason we will consider an approximate solution in two steps. In the regime $x\to 0$, the radial wave equation (\ref{ridotta}) can be approximated by
\begin{equation}
\frac{1}{x^2}\frac{{\rm d}}{{\rm d}x}\left(x^2\frac{{\rm d}f}{{\rm d}x}\right)-\sigma_0^2f(x)=0, \qquad x\to 0,
\label{ridottissima}
\end{equation}
while, in the regime $x\to 1$, the approximate form of (\ref{ridotta}) reads
\begin{equation}
\frac{1}{x^2}\frac{{\rm d}}{{\rm d}x}\left(x^2\frac{{\rm d}f}{{\rm d}x}\right)=0, \qquad x\to 1.
\label{piuridotta}
\end{equation}
Their respective solutions are
\begin{equation}
f_{0}^{+}(x)=\frac{1}{x}\cosh\left(\sigma_0 x\right),\quad f_{0}^{-}(x)=\frac{1}{x}\sinh\left(\sigma_0 x\right), \qquad x\to 0
\label{ddiia}
\end{equation}
and
\begin{equation}
f_1(x)=\frac{A}{x}+B, \qquad x\to 1.
\label{monete}
\end{equation}
As announced above, these eigenfunctions are real; by the discussion following Eq. (\ref{semplifikato}), they violate the cosmological principle only minimally. The functions $f_0^{\pm}$ must be joined smoothly to $f_1$ at some point $x_0\in[0,1]$; the joint function will be an approximate radial wavefunction for the vacuum state. 

Beginning with the hyperbolic sine first, let us consider the radial wavefunction
\begin{equation}
f(x)=\left\{\begin{array}{ll}
\sinh(\sigma_0 x)/x & \mbox{if $0\leq x\leq x_0$}\\
\\
A/x+ B & \mbox{if $x_0\leq x\leq 1$,}
\end{array}\right. 
\label{brenntwiesau}
\end{equation}
up to an overall normalisation factor $N(x_0)$. Dropping terms of order $\exp(-\sigma_0 x_0)$ and higher\footnote{This approximation is totally justified due to the sheer size of $\sigma_0=10^{123}$. For this approximation to break down one would have to go to a regime where $\sigma_0 x_0\simeq O(1)$ or, equivalently, $x_0\simeq 10^{-123}$. In turn, this would imply that the exponential part of the wavefunction should be strongly suppressed in favour of the term $A/x+B$. This, however, would be incompatible with the Hubble expansion.}, the matching conditions that $f$ and its derivative $f'$ be continuous at $x_0$ yield
\begin{equation}
A=-\frac{x_0\sigma_0}{2}\exp(\sigma_0 x_0),\qquad  B=\frac{\sigma_0}{2}\exp(\sigma_0 x_0),
\label{zichicchi}
\end{equation}
while for the normalisation factor $N(x_0)$ of $f$ we find
\begin{equation}
N(x_0)=\frac{\sqrt{12}\sigma_0^{-1}\exp(-\sigma_0 x_0)}{(1-x_0)^{3/2}}.
\label{eneequiscero}
\end{equation}
Eq. (\ref{eneequiscero}) is singular at $x_0=1$; this results from dropping subdominant terms. Had we dropped {\it no}\/ subdominant terms at all, then $N(x_0=1)$ would be perfectly regular. We can now compute the expectation value $\langle {\cal S}_g\rangle={\cal N}k_B\sigma_0\langle x^2\rangle$ as a function of the matching point $x_0$. We find
\begin{equation}
\langle x^2\rangle(x_0)=\langle f\vert x^2\vert f\rangle(x_0)=\frac{1}{10}\left(x_0^2+3x_0+6\right),
\label{ttilldd}
\end{equation}
which no longer exhibits any singularity since $\langle x^2\rangle(x_0=1)=1$. Some other values are
\begin{equation}
\langle x^2\rangle(x_0=0.9)=0.95,\quad \langle x^2\rangle (x_0=0.5)=0.77, \quad \langle x^2\rangle (x_0=0.1)=0.63.
\label{negljir}
\end{equation}
This result is easily interpreted: the Hubble potential drives an exponential expansion that causes the Universe to concentrate mostly around the boundary at $x=1$, even if the matching point $x_0$ is close to the origin. At the other end, when $x_0=1$, the corresponding entropy equals
\begin{equation}
\frac{\langle {\cal S}_g\rangle}{k_B}=\sigma_0\langle x^2\rangle= 10^{123},
\label{nopert}
\end{equation}
in complete agreement with the perturbative results of section \ref{hala}. In particular, the holographic bound continues to be saturated in this nonperturbative approach. The effect of having $x_0<1$ reduces this value somewhat, and the holographic bound is no longer saturated. However the reduction thus attained is negligible, far from the necessary $\sim10^{-19}$ that would be required to bring the entropy from the holographic bound $\sim 10^{123}$ down to its estimated value $\sim 10^{104}$ \cite{ASTROPH, FRAMPTON, PENROSE1, PENROSE2}.

One readily verifies that Eqs. (\ref{ttilldd}) and (\ref{nopert}) continue to hold if one replaces the hyperbolic sine with a hyperbolic cosine in the wavefunction (\ref{brenntwiesau}).

\subsection{Concluding remarks}\label{remaque}

In all three approaches considered here (perturbative using plane waves, perturbative using spherical waves, nonperturbative using approximate radial wavefunctions) we have abstained from applying boundary conditions to the wavefunction $\psi$. An obvious boundary condition to impose would be the vanishing of the wavefunction at $R_0$, the boundary surface of the Universe. Now requiring $\psi(R_0)=0$ would quantise the allowed energy levels. This represents no problem {\it per se}\/, but it creates some difficulties without actually improving our analysis. One expects the quantised energy levels to be so densely packed that, to all practical purposes, they will be indistiguishable from a continuous energy spectrum. On the other hand, the boundary condition $\psi(R_0)=0$ reduces the two linearly independent solutions of the Schroedinger equation to just one. For example, instead of the plane waves (\ref{wfkuu}) one would now have a sinusoidal wave vanishing at $R_0$, plus all of its higher harmonics. We do not gain much by this; but we {\it do}\/ lose some consistency, because sinusoidal waves (as opposed to the complex exponentials (\ref{wfkuu})) no longer satisfy the cosmological principle. Analogous arguments hold in the cases of the spherical waves (\ref{iggn}) and the hyperbolic functions (\ref{brenntwiesau}). Altogether, these considerations justify {\it not}\/ applying boundary conditions. 

The dimensionless parameters  $\lambda_0$  and $\sigma_0$ (Eqs. (\ref{teddy}) and (\ref{mundini})) carry opposite signs---they have to, as $\lambda_0$ is the energy of the groundstate of a negative potential, while $\sigma_0$ is its corresponding entropy. But they have the same absolute value. Given the physical constants at our disposal, $\sigma_0$ is the only (dimensionless) entropy and $\lambda_0$ is the only (dimensionless) energy that one can construct (up to dimensionless factors which our analysis cannot determine). So the equality $\sigma_0=-\lambda_0$ is inevitable. In turn, this equality reflects a physical property, namely: the equality of gravitational equipotential surfaces and isoentropic surfaces as dictated by the emergent spacetime scenario of ref. \cite{VERLINDE}, used here. 

One could turn the argument around and try to reason as follows. Starting from a knowledge of the actual entropy of the Universe $\sigma\sim10^{104}$, one derives the radial wavefunction describing this nonmaximally entropic Universe: one simply susbtitutes the dimensionless eigenvalue $\lambda=-\sigma=-10^{104}$ into the eigenfunctions (\ref{herri}), (\ref{tabernae}). Call the latter $\psi^{(j)}_{10^{104}}$ as in Eq. (\ref{karkon}).  The expectation value of $R^2$ in the states $\psi^{(j)}_{10^{104}}$ should give back the initial entropy $10^{104}$.

However, the above logic is flawed, because the eigenfunctions $\psi^{(j)}_{10^{104}}$ violate the cosmological principle substantially---and not just minimally, as argued in section \ref{vnasmn}. We can get an idea of the order of magnitude of this violation. The {\it radius}\/ $R$ of the Universe described by $\psi^{(j)}_{10^{104}}$  can be inferred from Eq. (\ref{ridotta}): write $\sigma=mH_0R^2/\hbar$, with $R$ replacing $R_0$, and solve for $R$. We find $R=8\times 10^{16}$ metres, a far cry from the actual radius of the Universe, $R_0=4\times 10^{26}$ metres.

The notion of the emergence of spacetime put forward in ref. \cite{VERLINDE} demands that, if the holographic bound is not to be saturated, then the quantum state of the Universe must be an excited state instead of the vacuum---it is only in a state of maximal entropy that minimal energy can be attained. Moreover, this must happen compatibly with the cosmological principle. Due to the limitations of our approach (the Newtonian approximation), the Universe described by our wavefunctions of sections \ref{hala} and \ref{npetrp} has more entropy than necessary. On the positive side,  the Universe described by our wavefunctions complies with the cosmological principle, with the holographic bound, and with the basic assumptions of emergent spacetime ({\it forces are entropy gradients}\/) put forth in ref. \cite{VERLINDE}.

\section{Discussion}\label{kkkksss}

In the nonrelativistic approximation, the cosmological fluid can be very conveniently described \`a la Madelung, by separating the wavefunction of the matter contents of the Universe into amplitude and phase. This observation opens the gate to the application of quantum mechanics in order to obtain estimates of thermodynamical quantities of the Universe, such as the gravitational entropy.

In section \ref{hala} we have carried out a perturbative computation. This perturbative analysis is based on a set of free wavefunctions, which one uses to evaluate the expectation value of  the Hubble potential.  The nonperturbative computation performed in section \ref{npetrp} is based on a set of interacting wavefunctions, obtained by solving the Schroedinger equation corresponding to the Hubble potential.

Both the perturbative and the nonperturbative analysis yield the same result: our estimates (\ref{estimado}) and (\ref{nopert}) saturate the upper bound established by the holographic principle \cite{BOUSSO}. Some estimates  \cite{ASTROPH, FRAMPTON, PENROSE1, PENROSE2} place $\langle {\cal S}_g\rangle/k_B$ at around $10^{104}$. While a somewhat lower value of our entropy would be clearly desirable, the fact is that the upper bound set by the holographic principle is respected by all our results. We are inclined to believe that the Newtonian approximation, applied throughout, is responsible for this saturation of the holographic bound, and that a fully relativistic treatment \cite{UPCOMING} will provide the necessary refinements that will reduce our entropy down to values better fitting with current estimates. Moreover, it is very rewarding to see the precise value of the holographic bound encoded in the wave equation as the parameter $\sigma_0$, see Eqs. (\ref{ridotta}) and (\ref{mundini}). This means that our crude model bears an element of truth.

Our analysis can be regarded as a quantum--mechanical application of the theory of emergent spacetime presented in the celebrated paper \cite{VERLINDE}. We have made decisive use of the property of {\it emergence}\/, both of classical spacetime and of quantum theory. As concerns spacetime, the emergent property is used when regarding gravitational equipotentials as isoentropic surfaces. Concerning quantum theory, emergence is used when regarding the wavefunction amplitude as the exponential of the (matter) entropy, as dictated by Boltzmann's principle. 

Admittedly, the assumptions made throughout automatically put black holes beyond our scope. Black holes are supposed to be the largest single contributors to the entropy budget of the Universe. Whether or not quantum mechanics as we know it remains applicable to black holes is of course a disputed question \cite{PENROSE1, PENROSE2}. This understood, we would like to point out that our estimate is based on a {\it dynamical}\/ model---a feature which, to the best of our knowledge, is entirely new.

\vskip0.5cm
\noindent
{\bf Acknowledgements} It is a great pleasure to thank Sarira Sahu, Daniele Tommasini and Joan V\'azquez Molina for interesting technical discussions. This research was supported by grant no. ENE2015-71333-R (Spain).

\end{document}